%% file: mnras_main.tex
\documentclass[fleqn,usenatbib,lineno]{mnras}

\usepackage[export]{adjustbox}
\usepackage{lineno}

\usepackage[T1]{fontenc}
\usepackage{lastpage}
\DeclareRobustCommand{\VAN}[3]{#2}
\let\VANthebibliography\thebibliography
\def\thebibliography{\DeclareRobustCommand{\VAN}[3]{##3}\VANthebibliography}

\usepackage{graphicx}	
\usepackage{amsmath}	
\usepackage{amssymb}	
\usepackage{subcaption}
\usepackage{caption}
\usepackage{graphicx}
\usepackage{float}
\usepackage{, bm}
\usepackage{pdflscape}

\usepackage{newtxtext,newtxmath}

\title[Classification of local ULIRGs and quasars]{Classification of local ultraluminous infrared galaxies and quasars with kernel  principal component analysis}

\author[Evangelos S. Papaefthymiou et al.]{\parbox{\linewidth}{
Evangelos S. Papaefthymiou, Ioannis Michos, Orestis Pavlou,Vicky Papadopoulou Lesta and Andreas Efstathiou\thanks{Corresponding author's e-mail: a.efstathiou@euc.ac.cy}}
\\
\\
School of Sciences, European University Cyprus, Diogenes Street, Engomi, 1516 Nicosia, Cyprus}

\date{\today}

\pubyear{2022}

\begin{document}
\label{firstpage}
\pagerange{\pageref{firstpage}--\pageref{lastpage}}
\maketitle

\begin{abstract}
We present a new diagnostic diagram for local ultraluminous infrared galaxies (ULIRGs) and quasars, analysing particularly the Spitzer Space Telescope's Infrared Spectrograph (IRS) spectra of 102 local ULIRGs and 37 Palomar Green quasars. Our diagram is based on a special non-linear mapping of these data, employing the Kernel Principal Component Analysis method. The novelty of this map lies in the fact that it distributes the galaxies under study on the surface of a well-defined ellipsoid, which, in turn, links basic concepts from geometry to physical properties of the galaxies. Particularly, we have found that the equatorial direction of the ellipsoid corresponds to the evolution of the power source of ULIRGs, starting from the pre-merger phase, moving through the starburst-dominated coalescing stage towards the active galactic nucleus (AGN)-dominated phase, and finally terminating with the post-merger quasar phase. On the other hand, the meridian directions distinguish deeply obscured power sources of the galaxies from unobscured ones. These observations have also been verified by comparison with simulated ULIRGs and quasars using radiative transfer models. The diagram correctly identifies unique galaxies with extreme features that lie distinctly away from the main distribution of the galaxies. Furthermore, special two-dimensional projections of the ellipsoid recover almost monotonic variations of the two main physical properties of the galaxies, the silicate and PAH features. This suggests that our diagram naturally extends the well-known Spoon diagram and it can serve as a diagnostic tool for existing and future infrared spectroscopic data, such as those provided by the James Webb Space Telescope.
\end{abstract}

\begin{keywords}
infrared: galaxies -- galaxies: evolution -- galaxies: Seyfert -- quasars: general -- methods: data analysis.
\end{keywords}



\vspace{-35pt}

\section{Introduction}

Local ultraluminous infrared galaxies (ULIRGs), with 1-1000$\mu m$ luminosities exceeding $10^{12} L_\odot$, since their discovery by the Infrared Astronomical Satellite (IRAS) in the 1980s \citep[][]{Neu1984,Soifer1984,Soifer1987}, have been studied extensively \citep[e.g.][]{Sanders1988,RowanEfstathiou93,Genzel1998,Farrah2003,imanishi07,vei09,Farrah2013,Efstathiou2022,Farrah2022}. It is now well understood that their infrared emission arises from a combination of star formation and active galactic nucleus (AGN) activity. Star formation dominates the far-infrared emission, whereas AGN can make a significant contribution and even dominate the emission of ULIRGs at near- and mid-infrared wavelengths \citep{Farrah2003,Vega08,Efstathiou2022}. However, disentangling the contributions of star formation and AGN activity to the luminosity still remains a major challenge. This is mainly due to the presence of dust that hides the energy sources of these galaxies. In particular, dust absorbs the optical and ultraviolet radiation emitted by the energy sources in ULIRGs and re-emits it as infrared radiation. Understanding local ULIRGs is of fundamental importance for interpreting submillimeter galaxies \citep[SMGs;][]{Hughes98,Barger1998,Casey2014} and other populations of galaxies, where extreme starburst and AGN activity occurred in the history of the Universe \citep[e.g.][]{Rowan-Robinson2018}.

The classical evolution scenario for ULIRGs suggests that they are evolutionary stages of a merger of two large gas-rich spiral galaxies \citep{SandersMirabel96,Perez2021}, which gradually lose their orbital energy and angular momentum to form a single galaxy. In this scenario, the end product of the evolution of ULIRGs is an important population of extragalactic objects known as quasars. Particularly, quasars represent the stage in the evolution of ULIRGs where the AGN has become the dominant energy source. In some sense, quasars are thought to be complementary objects to ULIRGs. However, there are studies which challenge this single evolutionary scenario and suggest the existence of multiple different evolutionary paths to be undertaken by ULIRGs \citep[see][and references therein]{Farrah2003}.  It is also worth noting that \citet{Harris2016} and \citet{Pitchford2016} found evidence for intense starbursts in luminous AGN, and also \citet{Kirkpatrick2020} recently detected a population of cold quasars which are not purely AGN objects, as they also seem to be accompanied by strong star formation activity.

The verification of the aforementioned scenarios for ULIRGs' evolution relies on the resolution of the main power source behind their infrared emission, and its relationship with their evolutionary stages. It is now well-established that the spectral energy distribution (SED) of infrared emission of ULIRGs displays many features that can be utilised to discern their nature. Particularly, the mid-infrared part of the spectrum has long been known to show a number of features due to interstellar dust, with excellent diagnostic power. These features arise due to the presence of polycyclic aromatic hydrocarbon (PAH) molecules and silicate dust grains. PAH features are generally regarded to give a strong indication of star formation from the general interstellar medium of the host galaxy \citep{Roche91,Genzel1998,efstathiou00,Peeters2002,imanishi07}.  Silicate dust can display absorption and emission features around 9.7 and 18$\mu m$. Absorption features are rather difficult to interpret, as they may indicate either a buried AGN \citep{imanishi07} or obscured star formation \citep{RowanEfstathiou93, imanishi07}. Silicate absorption can also be a result of the presence of a dusty torus around an AGN, when the AGN is viewed edge-on as predicted by radiative transfer models \citep{Pier1992,GranatoDanese1994,efstathiou95}. On the other hand, according to the same radiative transfer models for the torus, silicate emission features arise when AGN-dominated galaxies are viewed almost completely unobscured, i.e., the dusty torus is viewed face-on. Such interpretations are an essential element of the unified model of AGN \citep{Antonucci1993}.

Various methods have been used to decipher the dominant power source of ULIRGs based on the features of their infrared emission; such as SED fitting \citep{RowanEfstathiou93,Farrah2003,Marshall2007, Vega08,Efstathiou2014,Efstathiou2022}, development of diagnostic diagrams \citep{Genzel1998,Laurent2000,Spoon2007,Nardini2008,Nardini2010}, Graph theory \citep{Farrah2009}, Principal Component Analysis \citep{Wang,Hurley1} and Non-negative matrix factorization \citep{Hurley2}.

In this paper we propose a new classification diagram for ULIRGs and quasars, based on a nonlinear enhanced version of the well-known Principal Component Analysis (PCA), the so called {\it{Kernel Principal Component Analysis (KPCA)}}. This method has enabled us to classify the SED of the ULIRG sample of \citet{Farrah2009} to which we added the SED of 37 Palomar Green quasars \citep{Symeonidis2016}. We have managed to recover the four known optical classes of the galaxies, namely: H$_{\rm II}$ galaxies; low-ionization nuclear emission-line regions (LINERs); Seyfert 1 and Seyfert 2 galaxies \citep[e.g.,][]{Lisa_2006,imanishi07,Yuan_2010}. The novelty of our classification scheme lies in the fact that it distributes these classes on a well defined geometrical shape, i.e., an ellipsoid, whose equatorial and meridian directions correspond to physical characteristics of the galaxies. As it will be shown, this geometrically-oriented classification supports the temporal evolution scenario of the power source of ULIRGs and can also categorise galaxies based on the degree of obscuration of their dominant power source from dust, recovering subcategories of ULIRGs. To our knowledge, this geometrical interpretation is not something that has been considered before in observational astronomy.

Our results serve as an extension of previous efforts to classify ULIRGs. More specifically, our work naturally enhances the results of \citet{Wang} and \cite{Hurley1}, who had employed the classical PCA method. Furthermore, it generalizes the popular but heuristic diagram of \cite{Spoon2007}, recovering its axes as special projections of our diagram. Additionally, the classification scheme proposed in this paper improves the grouping in \cite{Farrah2009}, classifying unique galaxies which could not be incorporated in their graph based classification.

The paper is organized as follows:   Section \ref{sec:2} describes the observational and simulated data used in the analysis.  Section \ref{sec:3}  describes the Kernel PCA methodology applied on the data considered.   In Section \ref{sec:4} the  results of this work are presented and discussed. Finally,  Section  \ref{sec:5} presents the conclusions and some ideas for future work. Throughout this work, we assume \mbox{$H_0 = 70$\,km\,s$^{-1}$\,Mpc$^{-1}$}, \mbox{$\Omega = 1$}, and \mbox{$\Omega_{\Lambda} = 0.7$}.

\section{Data Description}
\label{sec:2}
\subsection{Observational data}

The sample of \citet{Farrah2009} consists of 102 ULIRGs with redshifts between $0<z<0.4$. The data were acquired by NASA's Spitzer Space Telescope's Infrared Spectrograph (IRS) instrument \citep{houck04, Werner2004}, which covers the wavelength range $5-35\mu m$ \citep{Lebouteiller2011}. The SEDs were extracted from the \textit{Combined Atlas of Sources with Spitzer IRS Spectra} (CASSIS) website, with ID:105 - \textit{Spectroscopic Study of Distant ULIRGs II} by James R. Houck \& Lee Armus of CASSIS version LR7, released in June 2015 by \cite{Lebouteiller2015}, for low resolution data with $60<R<125$. This dataset is comprised of 118 objects observed in Low-Resolution and 53 observed in High-Resolution. Due to the lack of any synchronous sky-background data for the High-Resolution set, we chose to study the Low-Resolution dataset, similarly to \cite{Farrah2009}.

The Palomar Green (PG) quasar SEDs we added to our sample were taken from \citet{Symeonidis2016}. We were able to extract IRS spectra from the CASSIS website for $58$ PG quasars. However, due to the fact that some of these quasars were only observed between $14-37\mu m$ - a region that excludes the range of interest studied in this work, namely PAH emission and silicate features - we had to limit our sample size to 42 quasars. Cross-referencing this sample with the ULIRG sample of \cite{Farrah2009}, we were able to add the aforementioned 37 PG quasars with redshifts between $0<z<0.2$ to the ULIRG sample, bringing our total sample size to 139 ULIRGs and PG quasars.

We developed our own Python code for extracting and pre-processing the necessary data from this dataset. Firstly, we extracted, cross-referenced and corrected the redshift values of each galaxy using the NASA Extragalactic Database (NED) \footnote{The NASA/IPAC Extragalactic Database (NED)
is operated by the Jet Propulsion Laboratory, California Institute of Technology,
under contract with the National Aeronautics and Space Administration.}. After filtering the SEDs for possible double flux values on certain wavelength data points found in the data\footnote{Double flux values were corrected via masking in order to extract a single value, based on the best-fit continuation of the signal.} and performing stitching for some observed jumps in the SED around $14\mu m$ (where the Short-Low and Long-Low modules of the instrument overlap), we calculated the rest wavelengths and flux densities, between $5.2-25 \mu m$, for use in our analysis.

 \subsection{Simulated Data}\label{Sim}

 We have also used simulated SEDs of ULIRGs and quasars computed with radiative transfer models, provided by \citet{Efstathiou2022} and references therein, in order to aid the interpretation of the distribution of ULIRGs and quasars on our proposed diagnostic diagram. Overall, we have generated $2000$ simulated galaxies - whose rest wavelengths cover a range from $5.2\mu m$ to $25\mu m$ - with resolution of $0.1\mu m$, to match the spectrum of the real galaxies. The models include the emission of an AGN torus according to \citet{efstathiou95} and \citet{efstathiou13}, starburst emission according to \cite{efstathiou00} and \cite{efstathiou09},  and emission from a spheroidal host \citep{Efstathiou2021}. These models are part of the CYGNUS\footnote{The models are publicly available at \url{https://arc.euc.ac.cy/cygnus-project-arc/}} collection of radiative transfer models. In these models, there exist overall 13 free parameters; 10 of them are defined in Table \ref{tab:models}, while the remaining 3 parameters are the scaling factors $f_{SB}$, $f_{AGN}$, $f_s$ that determine the luminosities of the starburst, AGN and spheroidal components respectively. To produce the simulated SEDs that are presented in Figure \ref{Comp} in Section \ref{sec:inter}, we randomly sample the 13 parameters of the CYGNUS model within the range given for the corresponding parameters in Table \ref{tab:models}. For  $f_{SB}$, $f_{AGN}$, and  $f_s$ we assume the ranges found in the SED fitting of the HERUS sample of local ULIRGs \citep{Efstathiou2022}.

\begin{table*}
	\centering
\caption{Parameters of the CYGNUS models used to produce the simulated SED of galaxies, symbols used, their assumed ranges and summary of other information about the models. There are 3 additional scaling parameters for the starburst, AGN and spheroidal models, $f_{SB}$,  $f_{AGN}$ and $f_s$, respectively.}
	\label{tab:models}
	\begin{tabular}{llll} 
		\hline
		Parameter &  Symbol & Range &  Comments\\
		\hline
                 &  &  & \\
{\bf Starburst}  &  &  & \\
                 &  &  \\
Initial optical depth of giant molecular clouds & $\tau_V$  &  50-250  &  \cite{efstathiou00}, \cite{efstathiou09} \\
Starburst star formation rate e-folding time       & $\tau_{*}$  & 10-30Myr  & Incorporates \citet{BruzualCharlot1993,BruzualCharlot2003}  \\
Starburst age      & $t_{*}$   &  5-35Myr &  metallicity=solar, Salpeter Initial Mass Function (IMF) \\
                  &            &  & Standard galactic dust mixture with PAHs \\
                   &           &         &    \\
{\bf Spheroidal Host}  &  &  &  \\
                 &  &  \\
Spheroidal star formation rate e-folding time      & $\tau^s$  &  0.125-8Gyr  & \cite{Efstathiou2003}, \cite{Efstathiou2021}  \\
Starlight intensity      & $\psi^s$ &  1-17 &  Incorporates \citet{BruzualCharlot1993,BruzualCharlot2003} \\
Optical depth     & $\tau_{v}^s$ & 0.1-15 &  metallicity=40\% of solar, Salpeter IMF\\
                  &            &  & Standard galactic dust mixture with PAHs \\
                  &            &  &  \\
{\bf AGN torus}  &  &    &  \\
                 &  &  &  \\
Torus equatorial UV optical depth   & $\tau_{uv}$  &  250-1450 &  Smooth tapered discs\\
Torus ratio of outer to inner radius & $r_2/r_1$ &  20-100 & \cite{efstathiou95}, \cite{efstathiou13} \\
Torus half-opening angle  & $\theta_o$  &  30-75\degr & Standard galactic dust mixture without PAHs\\
Torus inclination     & $\theta_i$  &  0-90\degr &  \\
                 &            & \\

                 \hline
	\end{tabular}
\end{table*}

  \vspace{-15pt}

\section{Methodology}
\label{sec:3}

In this work, we  perform an appropriate dimensionality reduction technique on the SED of the galaxies. We  implement the Kernel Principal Component Analysis (Kernel PCA) method in order to classify the galaxies based on their main features. Particularly, we  extract their reduced feature space, whose axes correspond to the principal components of their SED. In this space, galaxies are represented by dot-points and our findings suggest that they follow a well-defined geometric distribution. As we will present, the existence of such emerging geometrical shapes can reveal various physical characteristics of the galaxies.

 \subsection{Principal Component Analysis}
PCA is one of the most fundamental and physically intuitive dimensionality reduction methods. It has been widely used in various problems involving data clustering, motion and pattern recognition, image analysis etc. \citep[see references in][]{VidMaSas, ZakiMeira}. Importantly, its nonlinear extensions constitute one of the most prominent methods used by the Machine Learning community \citep{HofSchoSmo}.

Intuitively, PCA is a linear procedure that reduces the dimension of a data set. Specifically, it tries to find an appropriate low-dimensional orthogonal coordinate system for a given set of data, in such a way that its axes are indicating the direction of the principal variances of the data. The algebraic formulation of the PCA method is based on the notion of the eigenvalues and eigenvectors of a special symmetric and positive semi-definite matrix, as we will see below.

Assume that we are given a set of data consisting of $N$ elements, each of which encodes $D$ features. These elements can be represented algebraically by column vectors of dimension $D$, i.e., $\mathbf{x}_i\in\mathbb{R}^D$, for $i=1, \ldots , N$, so that our data set is a collection of points in a $D-$dimensional Euclidean space, which will be called the {\it{Input space}} of the data set. Eventually, using matrix notation, the whole data set can be written as a $D\times N$ matrix, which is denoted by $\mathbf{X}$.
For practical reasons, we also consider centered initial data by subtracting the average vector ${\bm{\mu}} = {\frac{1}{N}} \Sigma^N_{i=1}{\mathbf{x}_i}\in {\mathbb R}^{D}$ from the data and obtaining the new variables ${\overline{\mathbf{x}}}_i = {\mathbf{x}_i} - {\bm{\mu}}$ and the corresponding matrix ${\mathbf{\overline{X}}}$.

The PCA method tries to project optimally the data points onto a linear subspace of ${\mathbb{R}^D}$ of dimension $d$ (usually $d<<D$). In the general case of uncentered raw data, the PCA projects the data onto an affine subspace, i.e., the center of the data is not at the origin of the coordinate system. Optimal projection means maximization of the projected variance of the data \citep[for more details, see][]{VidMaSas, ZakiMeira}, which in turn reduces to the eigenvalue decomposition of the $D \times D$ matrix
\begin{equation}
 \mathbf{S}=\overline{\mathbf{X}} \, \overline{\mathbf{X}}^{\top}. \end{equation}
Note that if we divide $\mathbf{S}$ with the scalar $N-1$ we obtain the statistical notion of the  {\it{covariance matrix}}.
The principal directions of the variations of the data are precisely given by the eigenvectors $\mathbf{u}_j$ of the matrix $\mathbf{S}$, i.e.,
\begin{equation}
\mathbf{S}{\mathbf{u}}_{j} ={\lambda}_{j} {\mathbf{u}}_{j},
\quad j = 1, \ldots , D,
\end{equation}
where the eigenvalue ${\lambda}_{j}$ gives the corresponding projected variance. So, the total variance of the data is given by the sum of the eigenvalues $var(D)=\Sigma^{D}_{j=1} \lambda_i$. Note that these eigen-directions which maximize the projected variance are also the ones that minimize the mean square error (MSE) of the projection.

Since the matrix $\mathbf{S}$ is positive semi-definite, its eigenvalues are all non-negative numbers and can thus be arranged in descending order (i.e., $\lambda_1\geq \lambda_2 \geq . . . \geq \lambda_D \geq 0$). The corresponding eigenvectors are pairwise orthogonal, a fundamental property that, finally, provides linear independent principal directions. Therefore, the dimensional reduction of the problem is achieved by considering only the $d$-eigenvectors that correspond to the $d$-largest eigenvalues. Note that these principal directions are unique modulo rotations. Also, it can be shown that the mean square error of the dimensional reduction is given by the total variation of the data minus the projected variation of the data, i.e.,  $MSE=\Sigma^{D}_{i=1} \lambda_i-\Sigma^{d}_{i=1} \lambda_i$ \citep[see][]{ZakiMeira}.

It is worth noting that the data set in the new principal component basis lies on a $D$-dimensional ellipsoid whose semi-axes lengths are given by $\sqrt{\lambda_i}$, \citep[see][]{ZakiMeira}. Finally, we would like to emphasize that the principal directions are amenable to physical interpretation. An example of such an interpretation is the application of the PCA method on the genomes data of European citizens, where the major principal axes were found to coincide with the geographical coordinates of Europe's map \citep{Novetal}.
\subsection{Non-linear extensions of PCA}
In spite of its applicability in linearly structured data, the PCA method fails to describe data that have obvious non-linear structure. Therefore, various forms of non-linear PCA extensions have appeared in the literature over the last decades in order to deal with non-linear data \citep[see][]{VidMaSas,MaPosHer}. One of the most fundamental methods amongst them is the so called {\it{Kernel PCA}} method, which was introduced by \cite{SchTR}, implementing the well-known notion of {\it kernel} functions. The flexibility of this method, as we will explain below, makes it a suitable candidate for problems arising in physics.

In general, non-linear extensions of PCA are based on a procedure of embedding data into a higher-dimensional space, and it is in this new space where the PCA method is finally applied. This embedding can be visualized as an unprojection or unfolding of the non-linear structure of data into a space with more `room', where now the data displays a linear structure. This space will be referred to as the {\it{ Feature Space}} and will be denoted by $F$.

Mathematically, the method transforms the data from the Input space to the Feature space through a non-linear map which relates the feature variables to the input variables
\begin{gather*}
\phi : {\mathbb{R}}^{D}\longrightarrow F\\
\mathbf{x}_i \mapsto \phi(\mathbf{x}_i).
\end{gather*}
The Feature space $F$ can have an arbitrarily large, possibly infinite, dimension which will be denoted by $\widetilde{D}>>D$.
Now the new data matrix will be given in terms of the $\widetilde{D}\times\widetilde{D}$ matrix $\Phi$, formed by the columns of the centered transformed data  $\overline{\phi}(\mathbf{x}_i) = \phi(\mathbf{x}_i) - \phi({\bm{\mu}})$ as:
\begin{equation}\label{Stilda}
\widetilde{\mathbf{S}}= \mathbf{\Phi}\,\mathbf{\Phi}^{\top}.
\end{equation}
The principal directions in the Feature space $F$ correspond to the $N$ eigenvectors $\mathbf{v}_{1}, \mathbf{v}_{2}, \ldots , \mathbf{v}_{N}$ of $\widetilde{S}$, associated to the non-zero (not necessarily distinct) eigenvalues
${\widetilde{\lambda}}_{1}, \ldots , {\widetilde{\lambda}}_{N}$. The remaining $\widetilde{D} - N$ eigenvectors correspond to the zero eigenvalue. This is due to the fact that each of the vectors $\mathbf{v}_{1}, \mathbf{v}_{2}, \ldots , \mathbf{v}_{N}$ must lie in the span of $\phi({\mathbf{x}_{1}}), \ldots , \phi({\mathbf{x}_{N}})$ \citep[for more details see][]{VidMaSas}. This, in turn, suggests that there exist $N$-column vectors  $\mathbf{w}_{i}, \, i \in \{ 1, \ldots , N \}$, such that
\begin{equation} \label{wi}
   \mathbf{v}_{i} = \mathbf{\Phi}\mathbf{w}_{i}.
\end{equation}
Due to the increase of the dimension of the problem from $D$ to $\widetilde{D}$, the matrix $\mathbf{\widetilde{S}}$ can be potentially huge, which can lead to tremendous computational challenges. However, this hindrance can be bypassed utilizing an important result from Linear Algebra, which implies that the non-zero eigenvalues of the matrix $\widetilde{\mathbf{S}}$ are identical with the eigenvalues of its transpose \citep[p. 555]{Meyer}. Thus, eventually, one deals with the eigenvalue problem of the $N \times N$ matrix:
\begin{equation} \label{preGram}
\mathbf{G} := \widetilde{\mathbf{S}}^{\top} = \mathbf{\Phi^{\top}}\, \mathbf{\Phi}.
\end{equation}
It is not hard to show that the eigenvectors of $\mathbf{G}$ are precisely the $N$-vectors $\mathbf{w}_{i}$ of \eqref{wi} \citep[see][]{VidMaSas}, a relation that allows one to recover the principal directions (i.e., the vectors $\mathbf{v}_i$) of the problem. At this point, we would like to comment that we choose to normalize $\mathbf{w}_{i}$ according to ${{\lVert} {\mathbf{w}}_{i} {\rVert}}^{2} = {{\widetilde{\lambda}}_{i}}^{-1}$, so that the principal axes $\mathbf{v}_i$ become orthonormal, i.e., ${{\lVert} {\mathbf{v}}_{i} {\rVert}} =1$. Importantly, the computational cost of the problem depends solely on the number of the data and not on the dimension of their features.

A final issue with the non-linear PCA method is that, in general, $\phi$ is an unknown map. However, this can be resolved by the formulation of the Kernel PCA method, as it  will be described below.

\subsubsection{Kernel PCA}

 The $N \times N$ matrix $\mathbf{G}$ essentially defines similarity relations between data in the Feature space $F$. Specifically, this similarity is an {\em{inner product}} in $F$ and, therefore, it can be mathematically realized as a certain {\it{positive definite kernel function}} (i.e., a function which can be viewed as a matrix with positive eigenvalues) $K:  \mathbb{R}^D \times \mathbb{R}^D \longrightarrow \mathbb{R}$:

\begin{equation} \label{kerphi}
K(\mathbf{x}_i,\mathbf{x}_j):=\phi(\mathbf{x}_i)^{\top}\phi(\mathbf{x}_j),\, \, \mathbf{x}_{i},  \mathbf{x}_{j} \in\mathbb{R}^D.
\end{equation}
In terms of centered data, the formula for the corresponding kernel function, i.e.,  ${\overline{K}}(\mathbf{x}_i,\mathbf{x}_j) = \overline{\phi}{(\mathbf{x}_{i})}^{\top} \: \overline{\phi}(\mathbf{x}_{i})$, can be calculated by the matrix form expression  ${\overline{K}}={\mathbf{J}K}\mathbf{J}$, where $\mathbf{J}=\mathbf{I}-\frac{1}{N}\mathbf{1}$, with $\mathbf{I}$ being the identity matrix and $\mathbf{1}$ denoting the matrix whose elements are equal to one \citep{VidMaSas}.

In practice, since $\phi$ is unknown, in view of $\eqref{kerphi}$, we can use an a priori positive definite kernel, which, in turn,  will implicitly introduce  a corresponding map $\phi$ from the Input space to the Feature space $F$. This is due to a fundamental result from functional analysis, known as {\it{Mercer's Theorem}} \citep[see][]{VidMaSas,  BeChrRe}.

In the literature there is a variety of kernels which can be used to extract the type of non-linear structures that govern the physical problem at hand. The most commonly used  positive definite kernels are the following:
\begin{gather*}
\text{Linear}:\,\, K(\mathbf{x}_i,\mathbf{x}_j)=\mathbf{x}_i^{\top}\mathbf{x}_j+\alpha\\
\text{Gaussian}:\,\, K(\mathbf{x}_i,\mathbf{x}_j)=\exp{(-\gamma||\mathbf{x}_i-\mathbf{x}_j||^2_2)}\\
\text{Polynomial}:\,\, K(\mathbf{x}_i,\mathbf{x}_j)=(\beta \mathbf{x}_i^{\top}\mathbf{x}_j+\alpha)^{\delta},
\end{gather*}
where $\alpha,\beta,\gamma$ and $\delta$ are free parameters (with $\gamma > 0$), and whose values should be adjusted appropriately. Note that the linear kernel (the Euclidean dot product) recovers the linear PCA method, so the Kernel PCA naturally generalizes the PCA method. In our problem, the choice of the kernel and how its parameters should be tuned is presented in the subsequent section.

Finally, having chosen the kernel function $K$, now $\eqref{preGram}$ yields the following $N \times N$ {\it{Gramian}} matrix of the centered data:
\begin{equation} \label{Gram}
\mathbf{G}(K) = \begin{pmatrix} \overline{K}({\mathbf{x}_{1}}, {\mathbf{x}_{1}}) & \overline{K}({\mathbf{x}_{1}}, \mathbf{x}_{2})  & \cdots & \overline{K}({\mathbf{x}_{1}}, {\mathbf{x}_{N}}) \\
\overline{K}({\mathbf{x}_{2}}, {\mathbf{x}_{1}}) & \overline{K}({\mathbf{x}_{2}}, {\mathbf{x}_{2}}) & \cdots & \overline{K}({\mathbf{x}_{2}}, {\mathbf{x}_{N}}) \\
\vdots & \vdots & \cdots & \vdots \\
\overline{K}({\mathbf{x}_{N}}, {\mathbf{x}_{1}}) & \overline{K}({\mathbf{x}_{N}}, {\mathbf{x}_{2}}) & \cdots & \overline{K}({\mathbf{x}_{N}}, {\mathbf{x}_{N}})
\end{pmatrix}.
\end{equation}
The final step of the method states that for every data vector ${\mathbf{x}}_{i}$, its $k$-th non-linear principal component is given by the number:
\begin{equation} \label{Fcords}
 y_{ik} = {{\mathbf{w}}_{k}}^{\!\top}\, {Col}_{i}({\mathbf{G}}), \quad i,k = 1, \ldots , N,   \end{equation}
 where ${Col}_{i}({\mathbf{G}})$ denotes the $i$-th column of the matrix $\mathbf{G}$.

From a mathematical point of view it is worth noting that the above Gramian matrix ${\mathbf{G}}$ in the case of the Gaussian Kernel belongs to a special class of matrices, the so called {\it{totally positive}} (TP) matrices, i.e., matrices whose minors are all positive. TP matrices possess elegant properties (for instance, all their eigenvalues are positive and distinct) and are met in a wide range of physical problems \citep[see][]{Karlin}.

\subsubsection{Open issues in the Kernel PCA method}

The application of the Kernel PCA method imposes three open problems that need resolution. The first is the fact that the choice of the most appropriate kernel for the problem under investigation is a matter of experimentation. The second is related to the realization that the free parameters that are involved in various kernel functions are chosen on an ad-hoc basis. The third issue is the determination of the actual number of the principal components that should be considered in order to encapsulate the main variations of the data.

In the context of our problem, we overcome the first issue by choosing to apply the Gaussian kernel on our data. On the one hand, this is a widely used non-linear kernel function in machine learning. On the other hand, it naturally enhances the results of the linear PCA method already used in the classification of galaxies \citep[see][]{Hurley1,Wang}, as it implicitly incorporates linear and higher order polynomial functions. This can be realised by considering the Taylor expansion of the exponential function, i.e.,

\[\exp{(-\gamma||\mathbf{x}_i-\mathbf{x}_j||^2_2)}=1+\sum_{n=1}^{\infty} {\frac{{(-\gamma)}^{n}||\mathbf{x}_i-\mathbf{x}_j||^{2n}_2 }{n!}}.\]

The second and the third aforementioned issues, for the Gaussian kernel in particular, will be resolved  by using the notion of {\it{spectral gap}} from Matrix Analysis theory that takes into account the underlying structure of the data.

\vspace{-15pt}

\subsubsection{Spectral Gap}

In the literature, \citep[e.g., see][]{StZw}, it is commonly accepted that the appropriate reduction ($d$ out of $N$ principal variables) of the dimensionality of the Feature space is dictated by the largest gap in the eigenvalue spectrum of the Gramian matrix ${\mathbf{G}}$, i.e.,
\begin{equation}
\mathcal{\delta}({\mathbf{G}}) = \max_{1 \leq i \leq N-1} \bigl|{\widetilde{{\lambda}}}_{i+1} - {\widetilde{{\lambda}}}_{i}\bigr|.
\end{equation}
In particular, the dimension $d$ of the embedded space is considered to be equal to the smallest index $i^{*}+1$, where $\mathcal{\delta}({\mathbf{G}})$ occurs. Accordingly, the intrinsic dimension of the data is equal to $i^{*}$. The above view can be justified by recollecting the fact that the total variance of the data is given by the sum of their eigenvalues. By truncating the dimension of the data after the largest gap in the spectrum, we are taking into account the bulk of the total variance of the data.

Note that Gramian matrices, and dense matrices in general,  successfully detect the aforementioned intrinsic dimension due to a tellable spectral gap. By contrast, sparse feature matrices (e.g., the Laplacian matrix) do not yield such an estimate, since their eigenvalues are usually closely located \citep{Xiaoetal}.

Considering now the case of the Gaussian kernel, the Gramian matrix $G$, and hence the spectral gap ${\delta}(\mathbf{G})$, depends on the parameter $\gamma$. We intend to identify the particular value of $\gamma$ (if such a value exists uniquely) that maximizes ${\delta}(\mathbf{G})$. As we will show, it turns out, that for the data in hand, there exists a unique such value of $\gamma$.
\textbf{}
\begin{figure*}
\begin{subfigure}{.4\textwidth}
  \centering
 \includegraphics[width=1.1\textwidth]{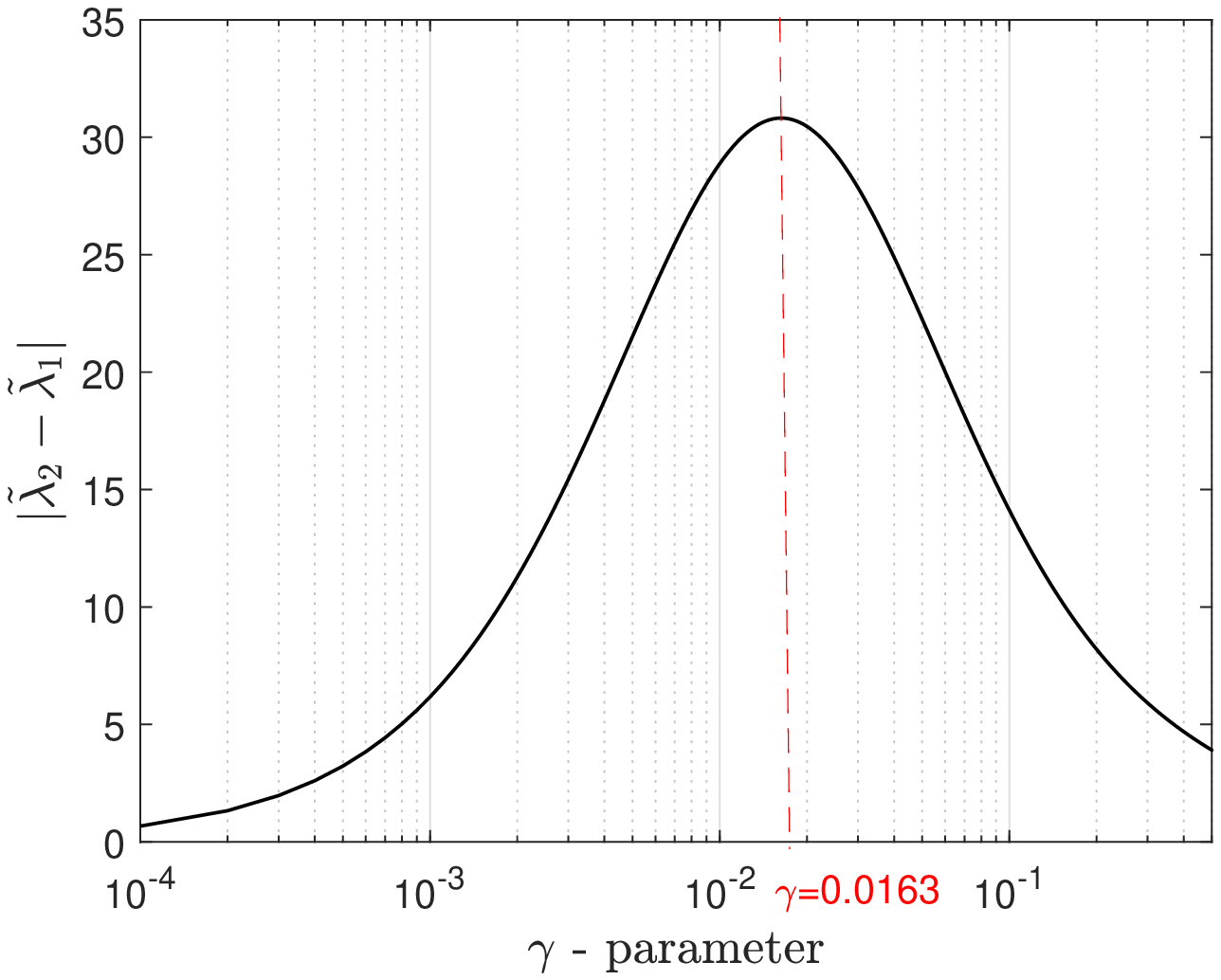}
\end{subfigure}
\begin{subfigure}{.4\textwidth}
 \centering
 \includegraphics[width=1.1\textwidth]{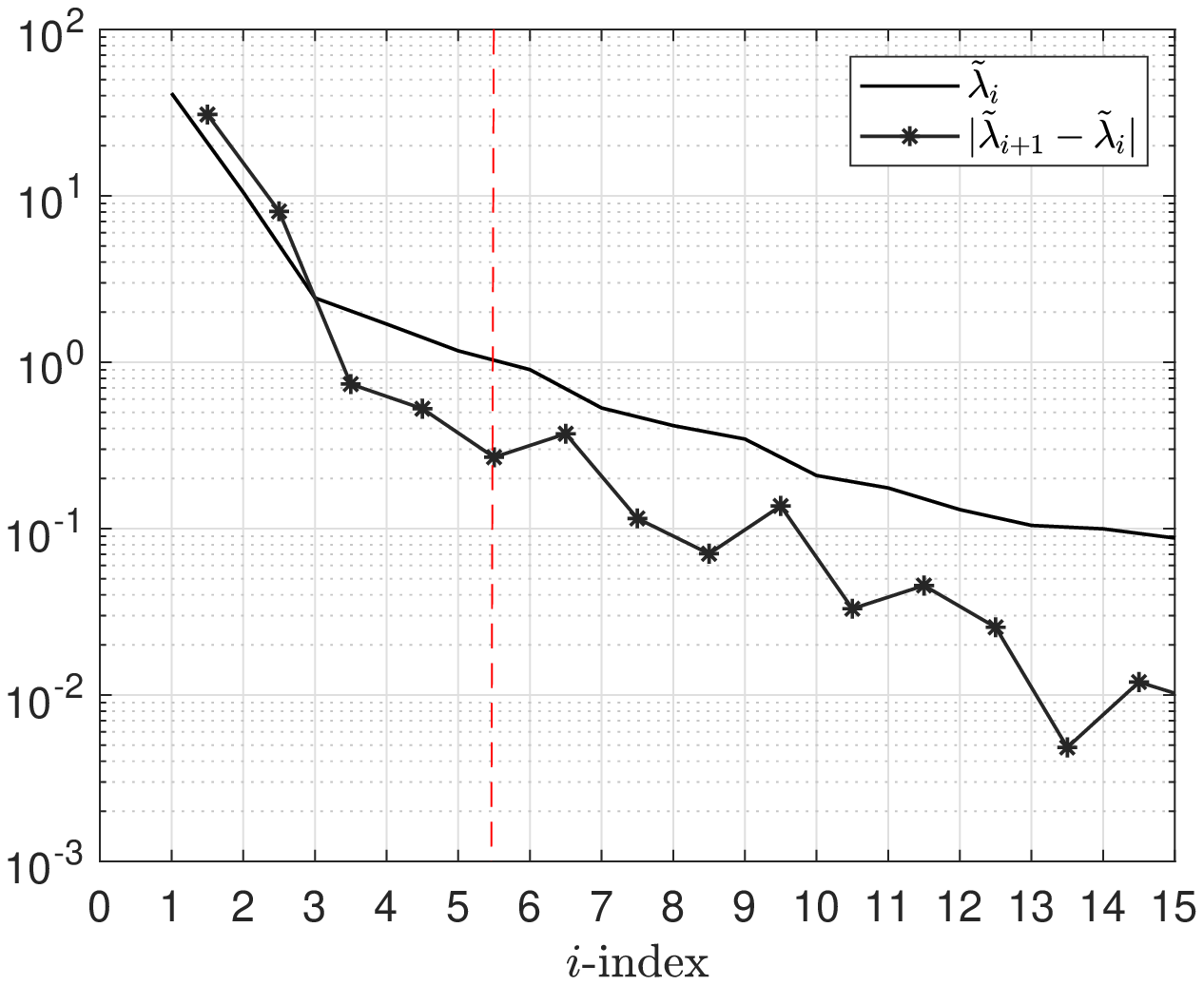}
\end{subfigure}
\caption{The left panel depicts the log-normal distribution of the absolute difference of the first two eigenvalues $|{\widetilde{\lambda}}_2 - {\widetilde{\lambda}}_{1}|$ of the Gramian matrix $(\ref{Gram})$ with respect to the parameter $\gamma$. The maximum of this difference occurs at $\gamma=0.0163$. The right panel depicts the eigenvalues (solid line) and the differences of successive eigenvalues (solid line with asterisks) of the Gramian matrix, when $\gamma=0.0163$. The difference of eigenvalues behaves monotonically up to the fifth successive difference $|\widetilde{\lambda}_6-\widetilde{\lambda}_5|$ (indicated by the red dashed line).}
\label{Lognormal}
\end{figure*}

 \vspace{-12pt}
 \section{Results and Discussion}
 \label{sec:4}

In this application the vectors $\mathbf{x}_i$ stand for the SED of the $139$ galaxies, i.e., $N=139$. Furthermore, the dimension of these vectors is $D=199$, since we consider SEDs in the rest-frame wavelength range 5.2-25$\,\mu m$, which is divided in steps of 0.1$\mu m$. The spectrum of each galaxy has been normalized with respect to its mean value as in \citet{Wang} and \citet{Hurley1}.

\begin{figure}
\begin{subfigure}{.5\textwidth}
  \centering
  \includegraphics[width=1.1\textwidth]{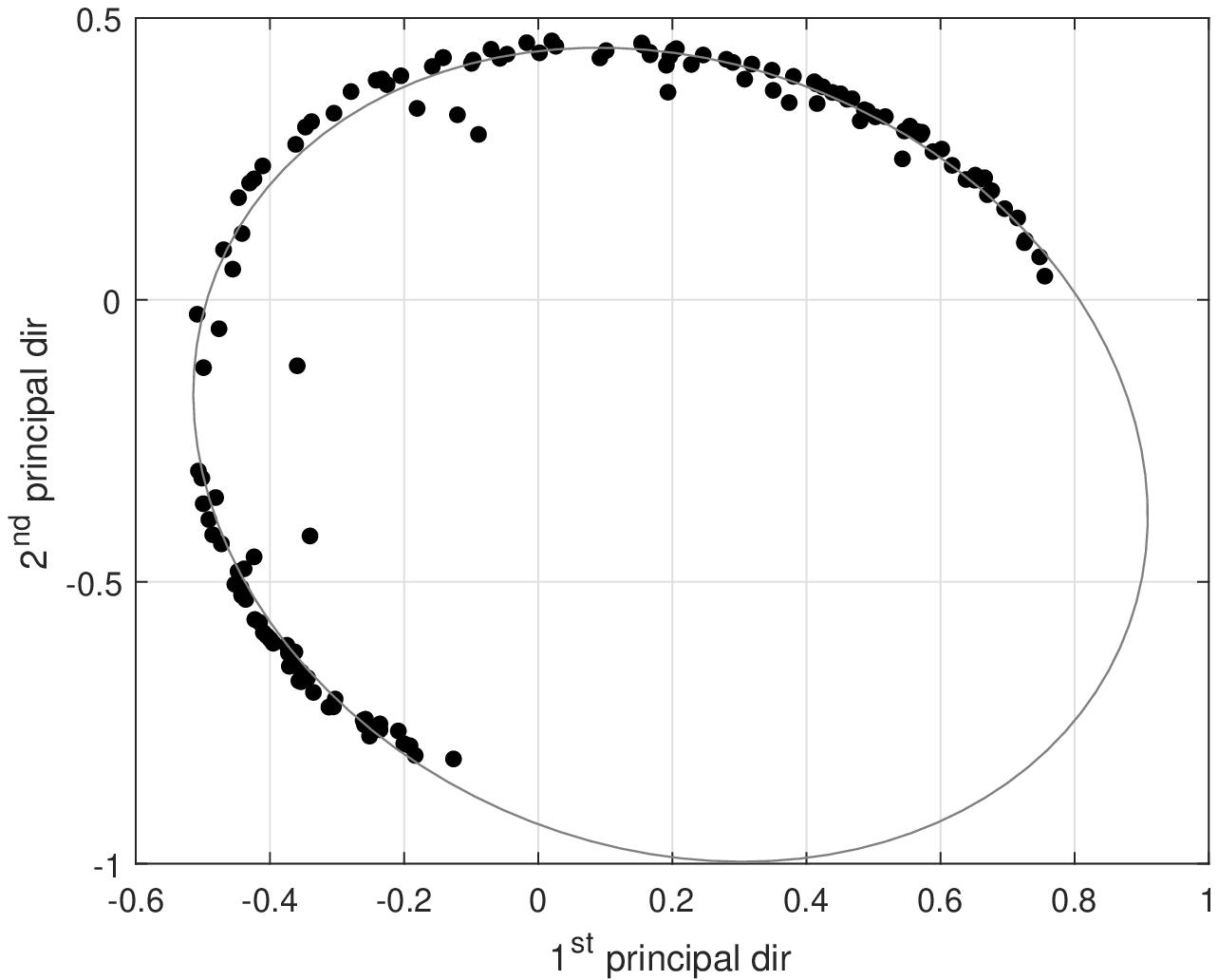}
\label{fig:ellipsis}
\end{subfigure}
\begin{subfigure}{.5\textwidth}
  \centering
  \includegraphics[width=1.1\textwidth]{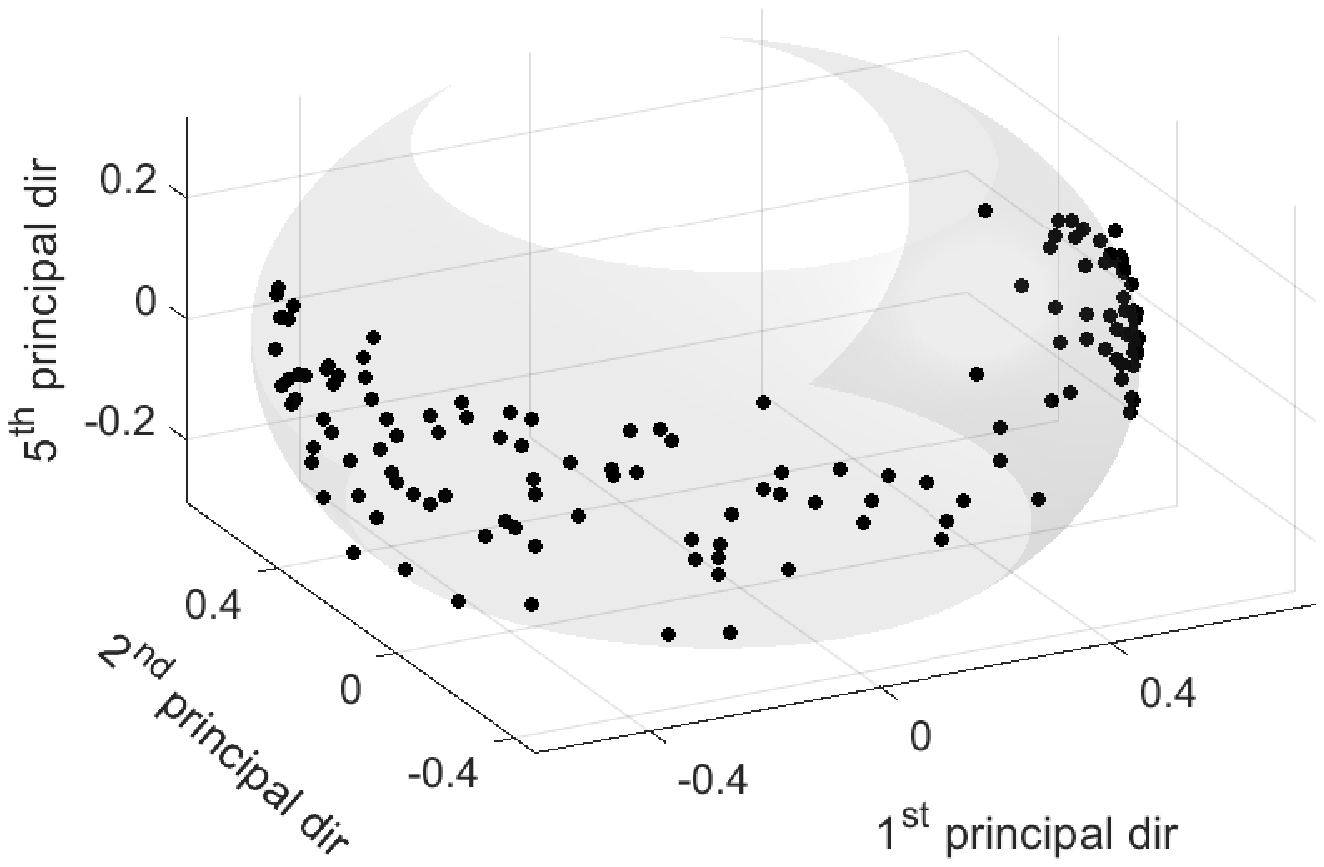}
\end{subfigure}
\caption{Distribution of the galaxies (black dots) in the Feature space produced by the Gaussian Kernel for $\gamma=0.0163$. The top panel displays the $2$-dimensional distribution of the galaxies on the $PC1$-$PC2$ plane; the bottom panel displays the $3$-dimensional distribution of the galaxies (black dots) in the $PC1$-$PC2$-$PC5$ space.}
\label{Manifold}
\end{figure}

\begin{figure*}
\centering
\includegraphics[width=0.7\textwidth]{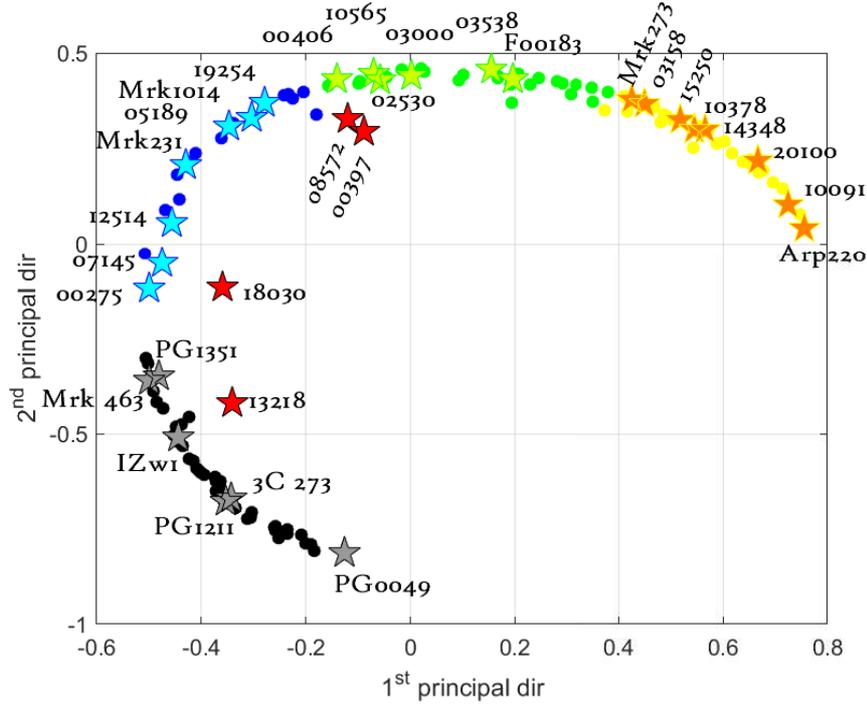}
\caption{Classification of ULIRGs and quasars based on their distribution on the $PC1$-$PC2$ plane. Four clusters, which are denoted with different colours, have been identified corresponding to the four main optical classes; yellow corresponds to LINERs; green to H$_{\rm II}$ galaxies; blue to Seyfert 2; and black to the Seyfert 1 class. Galaxies that lie away from the main distribution are not considered to belong to a particular class and are viewed as outliers, depicted with a red star. The rest of the shaded stars indicate noticeable examples of galaxies within each colored cluster.}\label{FeatureSpaceGaussianPCA}
\end{figure*}


\subsection{Learning the Manifold of the Distribution}

We now apply the Gaussian kernel in order to extract the underlying manifold of the distribution of the galaxies in their Feature Space \footnote{ For an open source code for the Kernel PCA method, see Masaki Kitayama (2021). MATLAB-Kernel-PCA (https://github.com/kitayama1234/MATLAB-Kernel-PCA), GitHub.}.

Initially, we search for the values of the parameter $\gamma$ of the Gaussian kernel that maximize the spectral gap ${\delta}(\mathbf{G})$ of the data set. We have considered a range of values within the interval $(0, 1)$. We first observe that for all the values of $\gamma$ in this interval, the spectral gap resides between the first and the second eigenvalue, i.e., $\delta(\mathbf{G}) = |{\widetilde{\lambda}}_2 - {\widetilde{\lambda}}_{1}|.$
Therefore, we can assume that the dominant variations of our data can be described in a $2$-dimensional space $(i^*+1 = 2)$, implying that the intrinsic dimension of the manifold of the distribution of the galaxies is equal to $1$.

We also observe that the relation between ${\delta}(\mathbf{G})$ and $\gamma$ follows a {\it log-normal} distribution, as it can be seen in the left panel of Figure \ref{Lognormal}. This leads to a {\em unique} value of $\gamma$ that maximizes the spectral gap, particularly $\gamma^{*} = 0.0163$, corresponding to the mean value of the log-normal distribution. The maximum spectral gap is then approximately equal to $30.82$.

\begin{figure}
\centering
\includegraphics[width=0.5\textwidth]{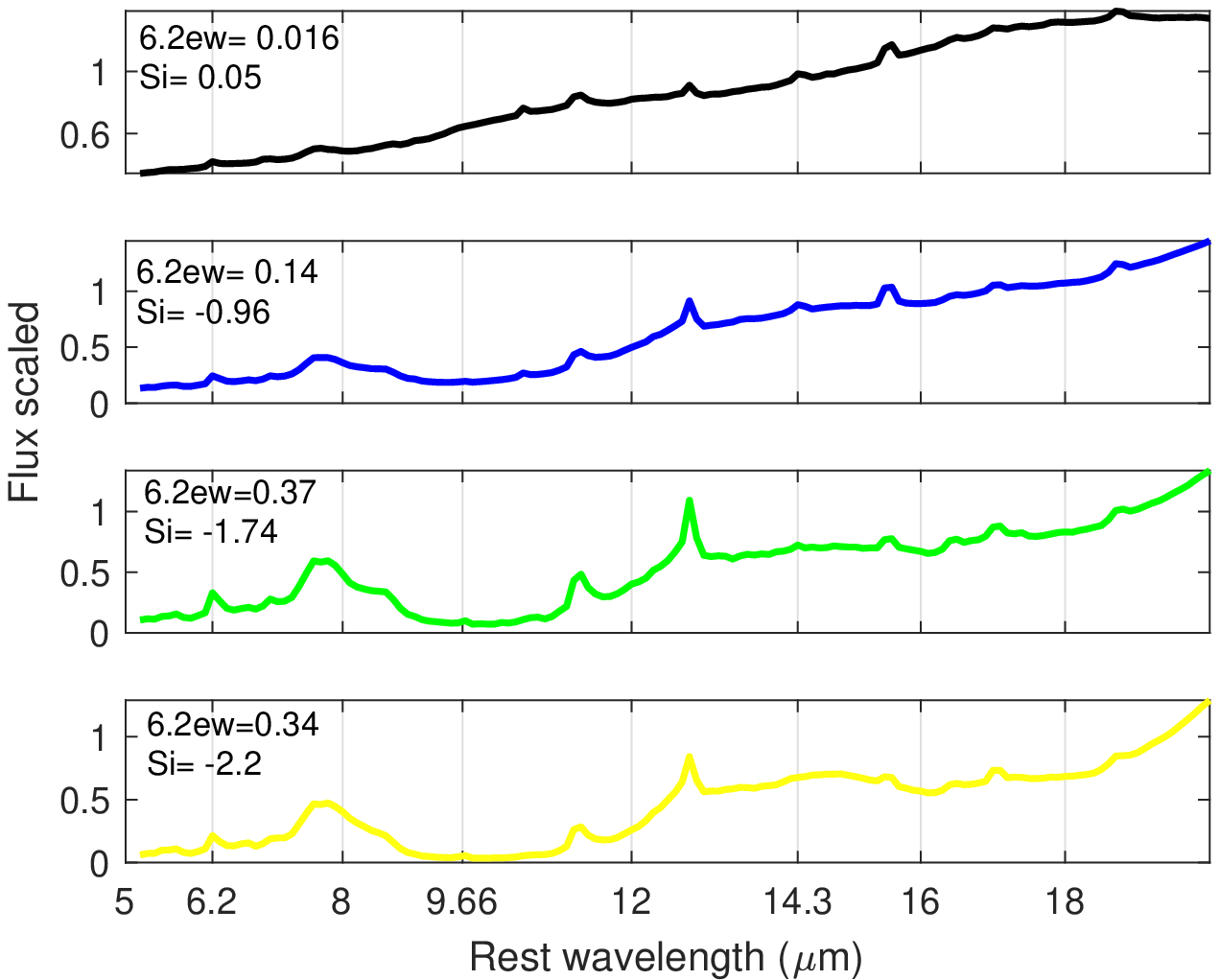}
\caption{Mean spectrum of the clusters. Each mean SED was created by averaging over the corresponding coloured clusters in Figure \ref{FeatureSpaceGaussianPCA}. The colours of the mean SEDs are identical with the respective cluster colours. In each mean SED, the corresponding 6.2$\mu m$ PAH equivalent width and 9.7$\mu m$ silicate strength (Si) values are displayed. The silicate strength is calculated as in Equation 1 in \citet{Spoon2007}}\label{MeanSp}
\end{figure}

As it can be observed at the top panel of Figure \ref{Manifold}, for the specific value ${\gamma}^{*} = 0.0163$, the distribution of the galaxies follows a well-defined geometrical curve \footnote{Curve fitting has been performed by utilising the open source code of Richard Brown (2021). fitellipse.m (https://www.mathworks.com/matlabcentral/fileexchange/15125-fitellipse-m), MATLAB Central File Exchange.}. In particular, it appears to be part of an ellipse, defined by the equation
$\displaystyle{\frac{(x-0.0834)^2}{0.7692^2}+\frac{(y-0.3276)^2}{0.6603^2}=1}$.
It is important to mention that even for small perturbations about the third decimal digit of the parameter $\gamma$, this tight geometrical shape is distorted, whereas in the case of values away from ${\gamma}^{*}$, the data are distributed unevenly. This suggests that the ellipse arises as a consequence of the fine tuning of the free parameter $\gamma$ in the Gaussian kernel. Such ellipsoidal geometrical structures are not a coincidence, they emerge as a mathematical by-product of the positive semi-definiteness of the Gaussian kernel, which theoretically distributes data on a $D$-dimensional hyper-ellipsoid. So, the natural question at this point is whether this emergent distribution are appropriate enough to capture the variations of the physical properties of the galaxies. We will try to answer this question in the next sections.

Based on the fact that there exist quite a few galaxies positioned distinctly away from the emergent ellipse, we proceed to explore also secondary variations of the distribution along additional directions. We will let again the spectral gap decide how many additional principal components should be considered. As shown on the right panel of Figure \ref{Lognormal}, the difference of the successive eigenvalues $|{\widetilde{\lambda}}_{i+1} - {\widetilde{\lambda}}_{i}|,\, i \in \{ 1, \ldots , N-1 \}$, behaves almost as a monotonically decreasing function, with the first exception at $|\widetilde{\lambda}_6-\widetilde{\lambda}_5|$, where one can spot a local minimum. After this point an oscillatory behavior appears, which can be seen as noise, since the magnitude of the corresponding eigenvalues is less than of order $10^{0}$. Computing the fraction $\displaystyle{\frac{\Sigma^{5}_{i=1} \widetilde{\lambda}_i}{\Sigma^{N}_{i=1} \widetilde{\lambda}_i}}$, we find that the first five components capture $93.4\%$ of the total variance of the data. Thus the first five principal components encapsulate most of the variation in our data. Note that this dimensionality reduction is also in agreement with the PCA analysis in \citep[]{Hurley1}.

The above discussion suggests that the data actually lives on a $5$-dimensional space.  For practical and visualization reasons we will focus our study on the $3$-dimensional projections of this space. Interestingly, by considering a third direction to be either one of the $3^{rd}$, $4^{th}$ or $5^{th}$ principal components, the actual distribution of the galaxies resides on a surface which is almost a {\it{spheroid}} (i.e., an ellipsoid of revolution). This can be seen for the case of the $5^{th}$ component in the bottom panel of Figure \ref{Manifold} \footnotemark[\value{footnote}]. Note that the ellipse depicted in the top panel of Figure \ref{Manifold} arises naturally as a projection of the spheroid on the $PC1$-$PC2$ plane.

\subsection{Interpretation of the results}
\label{sec:inter}
In this section, we attempt to interpret and classify the various types of ULIRGs and quasars under study, based on their distribution on the aforementioned emergent manifold.

We first focus on the $2$-dimensional distribution of the data on the $PC1$-$PC2$ plane (see top panel of Figure \ref{Manifold} and their corresponding labeling in Appendix  \ref{sec:A}). Motivated from the fact that there exist mainly $4$ optical classes for the galaxies under our study, i.e., H$_{\rm II}$ galaxies, low-ionization nuclear emission-line regions (LINERs), Seyfert 1 and Seyfert 2 galaxies \citep[e.g., see][]{Lisa_2006,imanishi07,Yuan_2010}, we search for $4$ clusters along the ellipse. To identify these, we implement the Kernel K-means method with $K=4$\footnote{The Kernel K-means method overcomes all the drawbacks of the classical K-means method when applied directly to the original input space of data that possess a nonlinear structure. It can efficiently detect non-convex clusters, as in the case of our problem, \citep[see][Section 13.2]{ZakiMeira}.}, which partitions the curve into 4 distinct segments.  These clusters can be seen in different colours in Figure \ref{FeatureSpaceGaussianPCA}.

A basic feature of the classification of Figure \ref{FeatureSpaceGaussianPCA} is the clear distinction of the galaxies based on their main power source. The yellow and green coloured galaxies represent the starburst-dominated sample of ULIRGs with a small contribution from an AGN, whereas the black and blue coloured ones are mostly consistent with Type 1 and Type 2 AGN respectively, characterized by a small starburst contribution. ULIRGs like the prototypical galaxy Arp~220, fit well within the starburst side of the curve, whereas all PG quasars are distributed near its opposite side. Examples of the latter, such as 3C~273, are bright quasars where the AGN torus is observed almost face-on (i.e., almost along the polar axis).

 The above interpretation can be supported by a thorough examination of  the mean SED of the four clusters shown in Figure \ref{MeanSp}. The green cluster matches an H$_{\rm II}$ profile, with strong PAH emission and silicate absorption features, whereas the yellow cluster's mean SED is more compatible with a LINER classification profile, displaying a slightly weaker PAH emission, but a deeper silicate absorption feature than H$_{\rm II}$ galaxies \citep{imanishi07}. The blue cluster's mean SED closely resembles a Type 2 AGN-dominated (Seyfert 2) galaxy SED, with minimal PAH emission and silicate absorption. The black cluster's mean SED is consistent with a Type 1 AGN-dominated (Seyfert 1) quasar, with weak PAHs and silicates in emission. The dissimilarity between the mean SED of the blue and black clusters strongly agrees with the AGN unification model \citep{Antonucci1993}, which suggests that the distinction between Type 1 and Type 2 AGN is mainly caused by the inclination of the observer's line of sight with respect to the polar axis: Type 1 AGN (black cluster) are observed face-on, whereas Type 2 AGN (blue cluster) are observed edge-on. Note also that Figure \ref{MeanSp} is in agreement with the PCA classification of \citep[]{Hurley1} presented in Figure 12 of their paper.

An interesting interpretation of Figure \ref{FeatureSpaceGaussianPCA}, in combination with Figure \ref{MeanSp}, is that it supports the ULIRGs' main power source evolutionary scenario \citep{Sanders1988,Rigopoulou1999, Veilleux2002, Veilleux2006, Spoon2007, Farrah2009}. This temporal evolution can be traced along the curve of Figure \ref{FeatureSpaceGaussianPCA}; beginning from the dust-dominated pre-merger phase of the yellow cluster; moving next to the starburst-dominated coalescence phase in the green cluster, where PAH features are highly prominent; then meeting the blue cluster where the AGN fraction increases and begins to dominate, diluting the PAH feature; and finally concluding with the luminous Type 1 AGN post-merger phase of the black cluster.

The above interpretation of temporal evolution of the galaxies along the curve, can also be justified by the following two arguments.

Firstly, we can consider the evolution of the neon fine structure line at $14.3 \mu m$ [Ne v] in Figure \ref{MeanSp}, which is known to be indicative of the presence of an AGN \citep{Sturm2002}. Comparing this feature between the mean SEDs, we can observe how it evolves from being almost completely absent in the yellow cluster to being very weak in the green cluster; the line evolves towards a clearly prominent feature in the blue and black clusters.

\begin{figure}
\centering
\includegraphics[width=0.5\textwidth]{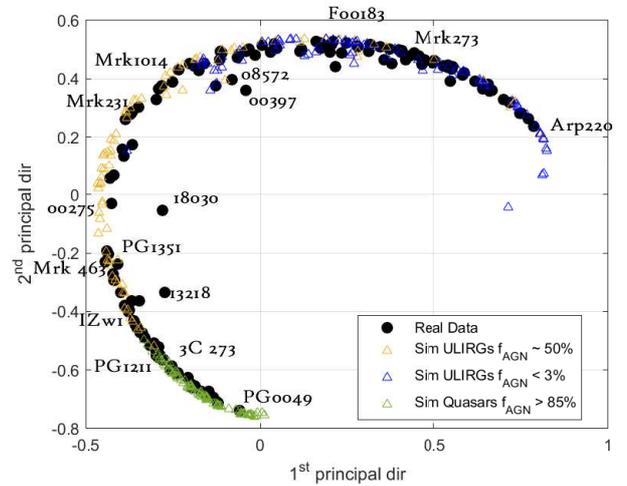}
\caption{Comparison of the PC1-PC2 distribution of the real (black circles) and simulated data of ULIRGs and quasars (coloured triangles). The yellow triangles are simulated galaxies generated assuming an AGN fraction of $\approx 50\%$, whereas for the blue and green triangles AGN fractions of $< 3\%$ and $> 85\%$ were assumed, respectively. For ULIRGs an edge-on inclination was assumed whereas for quasars it was assumed that the torus is viewed face-on.}
\label{Comp}
\end{figure}

Secondly, we have compared the distribution of the galaxies along the curve with the corresponding distribution of simulated galaxies. In particular, we have produced catalogs of simulated ULIRGs and quasars using the CYGNUS radiative transfer models employed by \cite{Efstathiou2022} to fit the SEDs of the HERUS ULIRGs, as this is described in Section \ref{Sim}. We have followed an identical procedure, as in the case of the real data, to calculate the corresponding $PC1$-$PC2$ Feature space. Figure \ref{Comp} displays this comparison and depicts that most of the simulated ULIRGs with AGN fraction less than $3\%$ coincide with the galaxies of the yellow and green clusters. On the other hand, most of the simulated ULIRGs with an AGN fraction of about $50\%$  reside on the blue cluster, and a small fraction of them covers a part of the black cluster. Finally, all the simulated quasars (which are generated assuming an AGN fraction greater or equal to $85\%$ and a face-on inclination) are concentrated in the black end of the curve, where real quasars also reside. We can therefore conclusively argue that the emergent distribution of the galaxies in their $2$-dimensional Feature space is compatible with the evolutionary scenario of ULIRGs' main power source.

If we now turn to the ULIRGs indicated with red stars in Figure \ref{FeatureSpaceGaussianPCA}, our diagram suggests that these are outliers of the overall sample, displaying certain unique characteristics that are responsible for positioning them outside the ellipse. Note that these galaxies were not included in the calculation of the mean SED of each cluster shown in Figure \ref{MeanSp}. Previous studies suggest that IRAS~08572+3915, which displays deep silicate absorption and minimal PAH emission \citep{Spoon2007}, is the most luminous ULIRG in the local ($z < 0.1$) universe and is thought to be an AGN-dominated galaxy with its torus viewed almost edge-on \citep{Efstathiou2014}; IRAS~18030+0705 displays the strongest PAH emission of all ULIRGs in the sample \citep{Spoon2007}; IRAS~13218+0552 is a known quasar with extreme outflows  \citep{Lipari2003}; and finally IRAS~00397-1312 is the most luminous galaxy in the sample \citep{Farrah2009}, showing deep silicate absorption and weak PAH features as in the case of IRAS~08572+3915.

\subsubsection{The distribution on the spheroid}

In further support of the above discussion, we also investigate the distribution of the galaxies on the spheroid that arises when we consider a third dimension in the Feature space. Our extensive investigations strongly suggest that the physical interpretation of the distribution of ULIRGs becomes more evident when the $5^{th}$ principal component is selected to be the third direction of the distribution. Therefore, our $3$-dimensional analysis will be solely carried out in the $PC1$-$PC2$-$PC5$ Feature space.

\begin{figure}
\centering
\includegraphics[width=0.5\textwidth]{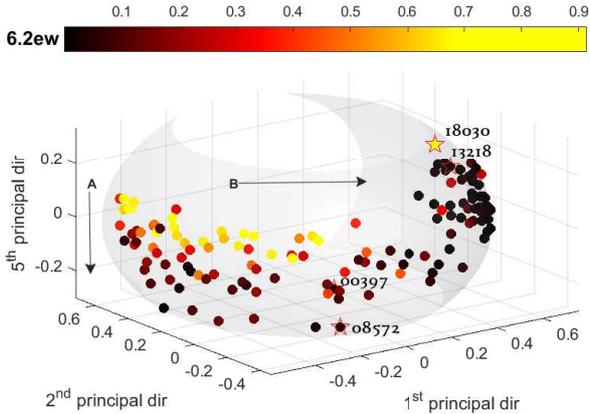}
\caption{6.2$\mu m$ PAH equivalent width variation across the $PC1-PC2-PC5$ dimensional spheroid. The variation across the spheroid shows the distribution of starburst-dominated ULIRGs (higher 6.2$\mu m$ PAH equivalent width) versus the distribution of AGN-dominated ULIRGs (lower 6.2$\mu m$ PAH equivalent width). Galaxies marked by asterisks are outliers and do not lie on the spheroid.}
\label{PAH}
\end{figure}

\begin{figure}
\centering
\includegraphics[width=0.5\textwidth]{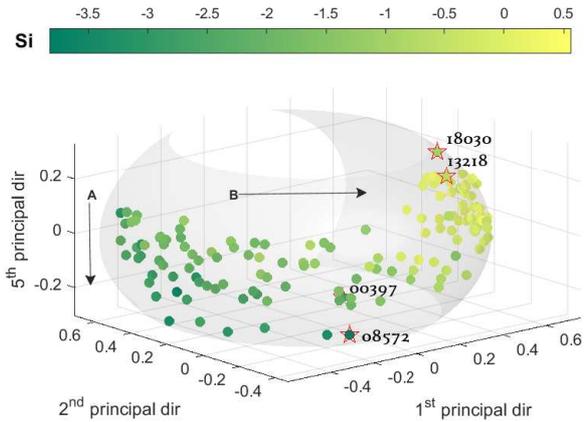}
\caption{9.7$\mu m$ silicate strength variation across the $PC1-PC2-PC5$ dimensional spheroid. ULIRGs with higher silicate strength (green) have higher dust obscuration, whereas ULIRGs with lower silicate absorption or emission (yellow) have lower dust obscuration and/or a more face-on torus inclination. Galaxies marked by asterisks are outliers and do not lie on the spheroid.}
\label{Sil}
\end{figure}

Figure \ref{PAH} shows the distribution of galaxies on the spheroid, coloured based on their 6.2$\mu m$ PAH equivalent width, which is considered to be directly proportional to starburst activity. On the other hand, Figure \ref{Sil} shows the same distribution but now coloured based on the 9.7$\mu m$ silicate strength, which is indicative of the ULIRGs' degree of obscuration but also torus inclination \citep{RowanEfstathiou2009}.

From Figures \ref{PAH} and \ref{Sil}, it can be observed that ULIRGs which reside on the left side of the spheroid show on average strong PAH emission, as well as dust absorption. These galaxies are starburst-dominated ULIRGs, identified previously in Figure \ref{FeatureSpaceGaussianPCA} as green and yellow clusters. What is of interest here is the fact that the third dimension can reveal the effect of dust obscuration on the PAH emission of starburst ULIRGs. In particular, by tracing the spheroid downwards (as indicated by arrow A in Figures \ref{PAH} \& \ref{Sil}), one can observe a considerable decline in PAH emission, and at the same time an increase in silicate absorption. Thus, we can argue that galaxies that reside close to the equator resemble the class of normal starburst galaxies with mixed dust/source geometry, displaying strong PAH emission independently of the silicate strength being present, as these have been identified by \cite{imanishi07} in Figure 1a. On the other hand, galaxies who reside lower (i.e., on the lower left part of the spheroid) are starbursts where the presence of dust obscures their PAH emissions, displaying almost no or weak PAHs and strong silicate absorption. These resemble to the category of galaxies as described in Figure 1d in \cite{imanishi07}, where a normal starburst nucleus is screened by foreground dust.

Transitioning now towards the right side of the spheroid (as arrow B points in Figures \ref{PAH} \& \ref{Sil}), PAH emission becomes less prominent on average, suggesting lower starburst activity and increased AGN fraction. Additionally, it can be observed that dust absorption decreases, turning eventually into emission. In conjunction with Figure \ref{Comp}, we can argue that part of the blue cluster (i.e., galaxies that reside closer to the green cluster) consists of buried AGN/starburst composite galaxies, where the observed spectrum is a superposition of PAH emission from starburst and emission from AGN. This resembles to the class of galaxies that correspond to figure 1c in \cite{imanishi07}. The remaining part of the blue cluster (those galaxies that are closer to the black cluster) consists of galaxies displaying purely buried AGN features corresponding to Figure 1b \cite{imanishi07}. Note that at this part of the spheroid, silicate features transition from absorption to emission, and obscuration can be related to the inclination view of the AGN torus, which conceals the AGN source. This transition can be located at the boundary between the blue and black clusters. Finally, galaxies at the right most part of the spheroid (black cluster), exhibit minimal PAH emission and silicate features are in emission, indicating the presence of quasars, in which the torus is viewed face-on.

Based on the above observations, we can conclude that our  $3$-dimensional distribution of the galaxies is able not only to separate AGN-dominated ULIRGs from starburst-dominated ones, but also can distinguish sub-classes among them, based on the degree of obscuration of their dominant power source from dust. At this point, we would like to comment that an interesting question is whether various geodesic paths on this manifold correspond to possible secondary physical evolutionary scenarios of the galaxies. This is a research direction we intend to pursue in a future work.

\begin{figure}
\begin{subfigure}{.5\textwidth}
  \centering
  \includegraphics[width=0.8\textwidth]{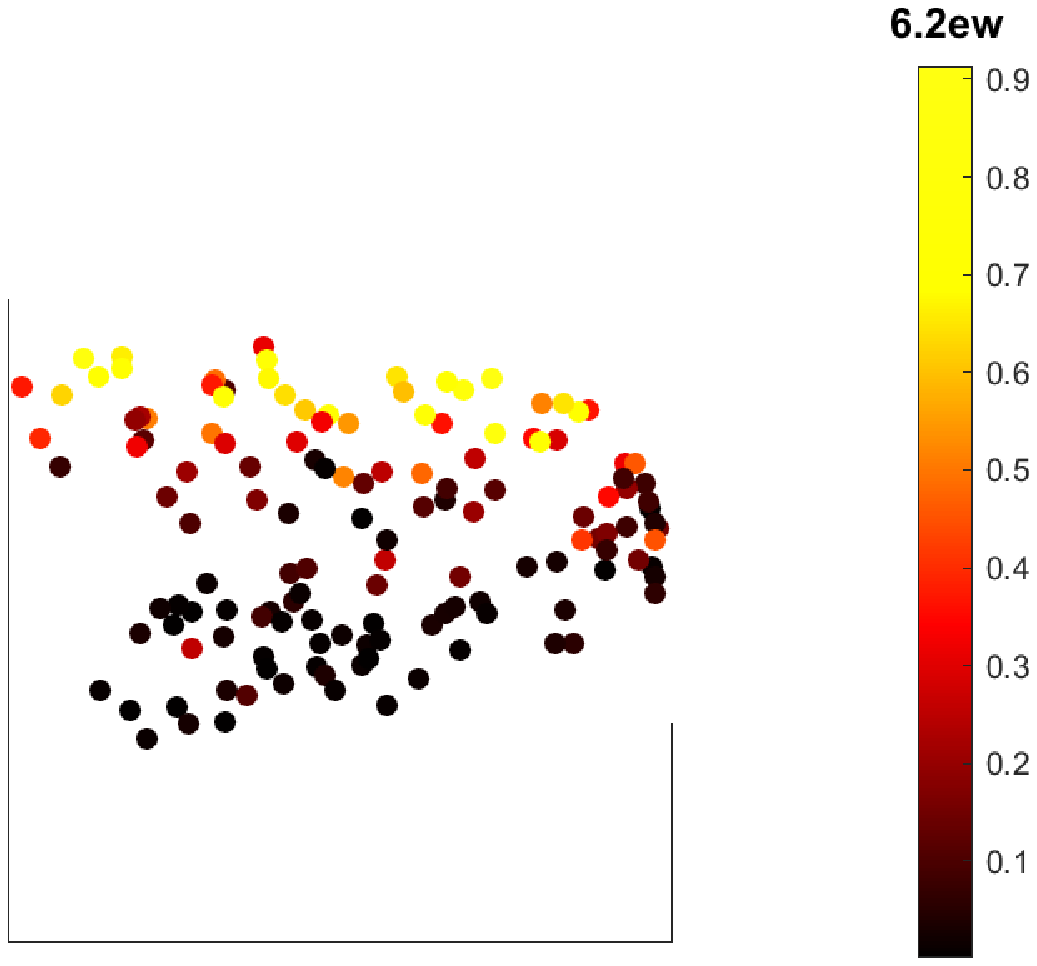}  
  \label{fig:sub-first}
\end{subfigure}
\begin{subfigure}{.5\textwidth}
  \centering
  \includegraphics[width=0.8\textwidth]{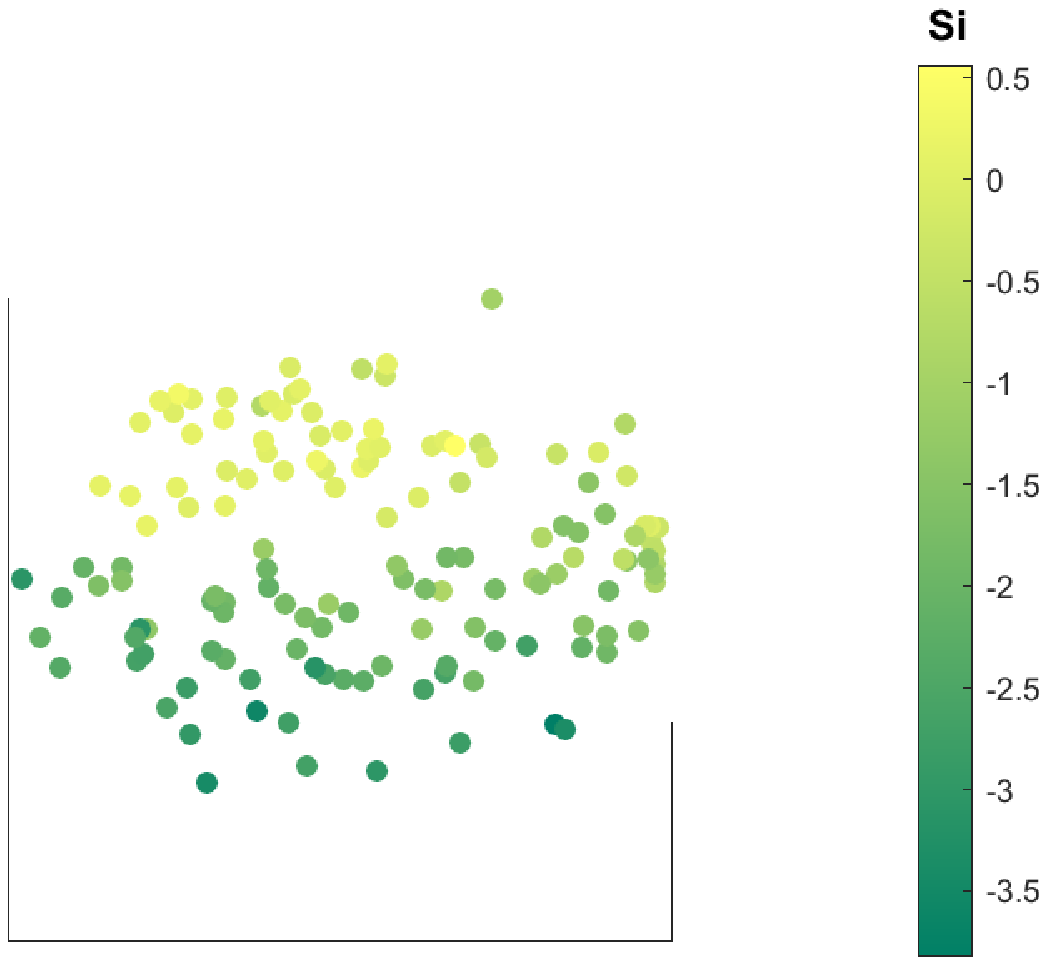}
\end{subfigure}
\caption{ Special projections of the spheroid on the $PC2-PC5$ plane. Top: 2-dimensional projection colour-coded with the 6.2$\mu m$ PAH equivalent width variation. Bottom: 2-dimensional projection colour-coded with the 9.7$\mu m$ silicate strength variation. These projections show how a monotonic variation of the two main physical characteristics of the galaxies can be recovered by projecting the spheroid onto the $PC2-PC5$ plane.}
\label{Proj}
\end{figure}

Another interesting point regarding the distribution of galaxies on the spheroid is the following. If we consider two particular projections of the spheroid on the $PC2$-$PC5$ plane, we can recover the physical axes of the galaxies that correspond to the PAH and  silicate features of the SEDs. Particularly, after rotating the spheroid clockwise through an angle of $15\degr$ with respect to the $PC2$ axis, an almost monotonic increment of the PAH values of the galaxies, starting from the bottom to the top of the projection, emerges as the top panel of Figure \ref{Proj} depicts. On the other hand, a counterclockwise rotation, again through an angle of $15\degr$, an almost monotonic distribution of the silicate strength of the galaxies can be revealed, as the bottom panel of Figure \ref{Proj} displays. Interestingly, through these special projections we have indirectly recovered the axial directions of the well-known Spoon diagram \citep[][]{Spoon2007}.

\begin{figure}
\centering
\includegraphics[width=0.5\textwidth]{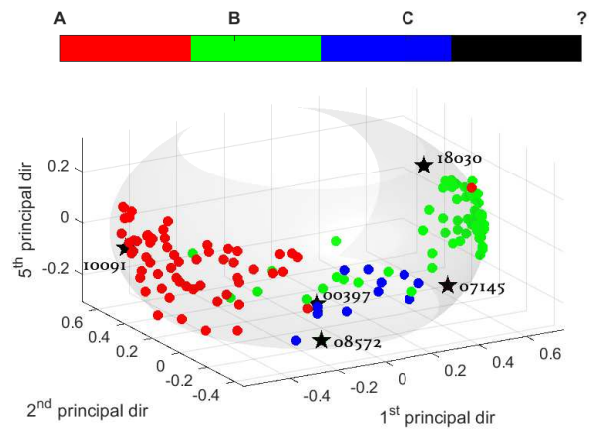}
\caption{Clusters colour-coded using the \citet{Farrah2009} cluster classification. Group A, containing starburst-dominated ULIRGs, is shown in red. Group B, which mostly consists of AGN-dominated ULIRGs, is in green. Group C, in blue, most likely consists of ULIRGs displaying a mixture of starburst and AGN features in a transitional phase between the two stages. Black stars correspond to unique galaxies that were not identified by \citet{Farrah2009}.}
\label{FarrahClus}
\end{figure}

At this point we would also like to compare our results with the findings of \cite{Farrah2009}. Figure \ref{FarrahClus} displays our 3-dimensional spheroid distribution, but now the colouring corresponds to the groups identified by \cite{Farrah2009}, who used a graph theoretical analysis approach. Group A galaxies were identified as ULIRGs displaying more prominent starburst features, whereas Group B ULIRGs display features more consistent with AGN-dominated and quasar-like galaxies. Our diagram supports these findings, where Group A represents the pre-merger and coalescence evolutionary stages (our yellow and green clusters of Figure \ref{FeatureSpaceGaussianPCA}) and Group B represents the post-merger phase of the evolutionary scenario of the galaxies (our blue and black clusters). However, our first interpretation of our diagram does not support the assumption that ULIRGs in group C, as identified by \cite{Farrah2009}, form a separate group that potentially represents a different evolutionary stage. Instead, our diagram indicates that this group belongs to a transitional phase between the coalescence and the post-merger stage. It appears to be more similar to AGN-dominated galaxies (they belong to our blue cluster), most likely consisting of ULIRGs displaying a mixture of starburst and AGN features. On the other hand, we would like to comment that such alternative evolutionary scenarios may coexist to the primary one, and the identification of these in our case demands a detailed exploration of various geodesic paths of our diagram; this is left for a future work. Finally, unique objects displayed by a black star in Figure \ref{FarrahClus} that \cite{Farrah2009} were not able to  classify due to limitations of their method, can now be positioned in the 3-dimensional Feature space.

To conclude our study we turn our attention towards the outlier ULIRGs that have been identified in the 2-dimensional case (i.e., the $PC1-PC2$ plane) of our analysis. These galaxies are also the only galaxies that do not lie on the spheroid of the 3D Feature space. Note that this deviation might not be discernible in Figures \ref{PAH} and \ref{Sil}. However their position on the 3D space can give indications about the origin of their unusual properties. Specifically, ULIRG IRAS~08572+3915 at $z=0.058$ is located very low at the vertical (PC5) axis of the 3D Feature space and at the same time is closer to galaxies identified to display AGN/starburst composite spectrum. This suggests that it can posses composite AGN/starburst power source, but dust obscures it in a drastic way. However, deep obscuration by dust may not be the only factor that makes IRAS~08572+3915 an outlier because Arp~220 which also shows evidence of deep obscuration is not an outlier. Similarly, IRAS~00397-1312, which is the most luminous ULIRG in our sample, is also placed low on the PC5 axis, suggesting again a deeply obscured power source. On the other hand IRAS~18030+0705 - an extraordinary object with highly peaked spectrum - occupies the highest point on the vertical (PC5) axis of our diagram, indicating that its dust content does not affect its main features. Our diagram suggests that IRAS~18030+0705 resembles more an AGN-dominated galaxy, since it occupies the upper-right part of the spheroid, but it simultaneously displays other extreme features, such as extreme PAH values.

\section{Conclusions and Future Work}
\label{sec:5}

In this work, we have presented a new classification diagram for the SED of the ULIRG sample of \citet{Farrah2009} along with the SED of 37 Palomar Green quasars \citep{Symeonidis2016}. This diagram has been constructed by implementing a nonlinear unsupervised machine learning technique, the Kernel PCA method, taking into account the whole mid-infrared spectrum of the galaxies. This implementation introduces a new perspective which uses ideas from differential geometry to classify ULIRGs and quasars; it uniquely positions the galaxies on an ellipsoid, whose intrinsic directions correspond to specific attributes of the galaxies. Particularly, the equator direction of the ellipsoid corresponds to the evolution of the galaxies, sequentially transitioning through the four main optical classes: LINERs, H$_{\rm II}$, Seyfert 2 and Seyfert 1. This transitioning distribution along the equator of the ellipsoid supports the scenario for the evolution of the power source of ULIRGs. On the other hand, the meridian direction of the ellipsoid indicates the strength of the concealment of the main power source of galaxies due to the existence of dust; one of the main hindrances in understanding the various physical processes and power sources that are present in ULIRGs. Furthermore, we should point out that our method has successfully identified unique objects with extreme features.

Finally, we would like to discuss a future direction of the proposed method on the analysis of the SED of unidentified ULIRGs. Our diagram can be utilized as a supervised machine learning tool, by considering it as a cartogram, whose coordinates correspond to specific physical attributes. By positioning unknown galaxies in this map one can uncover their physical characteristics through comparisons with the coordinates of known galaxies. This can be appropriate for analysing larger samples of galaxies observed either with Spitzer IRS, or with the forthcoming data from the James Webb Space Telescope.

\vspace{-15pt}
\section*{Acknowledgements}

The authors acknowledge support from the project EXCELLENCE/1216/0207/ GRATOS funded by the Research \& Innovation Foundation in Cyprus.

\vspace{-15pt}
\section*{Data Availability Statement}

The observational data underlying this article are available from the \textit{Combined Atlas of Sources with Spitzer IRS Spectra} (CASSIS) website. The simulated data used in this article were generated using the CYGNUS radiative transfer models available at https://arc.euc.ac.cy/cygnus-project-arc/.

\vspace{-15pt}

\bibliographystyle{mnras}
\bibliography{ReferencesBib} 

\newpage
\appendix

\input{AppendixA.tex}

 \label{lastpage}
\end{document}

%% file: AppendixA.tex
\section{Labeling and Classification of the sample}
\label{sec:A}

\begin{figure*}
\centering
\includegraphics[scale=0.38,left]{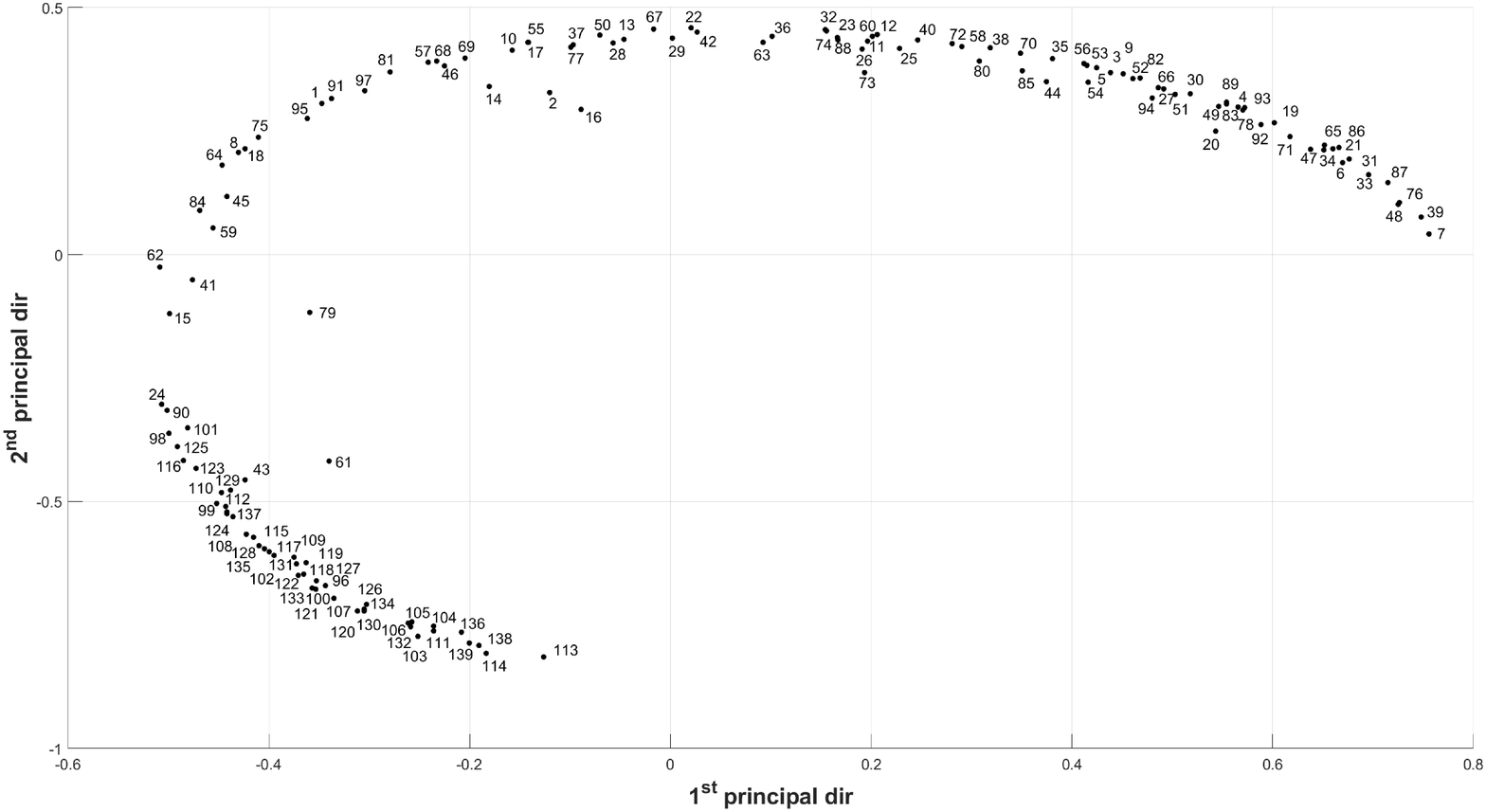}
\caption{Distribution of the sample of galaxies studied in this work on the $PC1$-$PC2$ plane along with their corresponding IDs.}
\label{AppendixA_figure1}
\end{figure*}

Figure \ref{AppendixA_figure1} depicts the IDs of the galaxies presented in Figure \ref{FeatureSpaceGaussianPCA}. Table \ref{tab:ID} gives the names of the sample of galaxies studied in this work with their corresponding IDs, and their classification based on the four clusters that have been identified in Figure \ref{FeatureSpaceGaussianPCA}.   
 
{\scriptsize
\begin{table}
\caption{IDs and Classification of the ULIRGs and quasars in our sample}
\label{tab:ID}
\begin{tabular}{lcc} \hline
ID & Galaxy & Classification  \\
\hline
1 & IRAS~05189-2524 & S2 \\ 
2 & IRAS~08572+3915 & S2\\
3 & IRAS~12112+0305 & LINER\\ 
4 & IRAS~14348-1447 & LINER \\
5 & IRAS~15250+3609 & LINER\\
6 & IRAS~22491-1808 & LINER\\
7 & Arp~220 & LINER \\
8 & Mrk~231 & S2\\
9 & Mrk~273 & LINER \\
10 & UGC~5101 & H$_{\rm II}$\\
11 & IRAS~F001837111 & H$_{\rm II}$\\
12 & IRAS~00188-0856 & H$_{\rm II}$\\
13 & IRAS~00199-7426 & H$_{\rm II}$\\
14 & IRAS~00275-0044 & S2\\
15 & IRAS~00275-2859 & S2\\
16 & IRAS~00397-1312 & S2 \\
17 & IRAS~00406-3127 & H$_{\rm II}$\\
18 & IRAS~01003-2238 & S2\\
19 & IRAS~01199-2307 & LINER\\
20 & IRAS~01298-0744 & LINER\\
21 & IRAS~01355-1814 & LINER\\
22 & IRAS~01388-4618 & H$_{\rm II}$ \\
23 & IRAS~01494-1845 & H$_{\rm II}$\\
24 & IRAS~02054+0835 & S1\\
25 & IRAS~02113-2937 & H$_{\rm II}$\\
26 & IRAS~02115+0226 & H$_{\rm II}$\\
27 & IRAS~02455-2220 & LINER \\
28 & IRAS~02530+0211 & H$_{\rm II}$\\
29 & IRAS~03000-2719 & H$_{\rm II}$\\
30 & IRAS~03158+4227 & LINER \\
31 & IRAS~03521+0028 & LINER \\
32 & IRAS~03538-6432 & H$_{\rm II}$ \\
33 & IRAS~04114-5117 &  LIN\rm ER \\
34 & IRAS~04313-1649 & LINER \\
35 & IRAS~04384-4848 & LINER \\
36 & IRAS~06009-7716 & H$_{\rm II}$\\
37 & IRAS~06035-7102 & H$_{\rm II}$ \\
38 & IRAS~06206-6315 & H$_{\rm II}$ \\
39 & IRAS~06301-7934  & LINER  \\
40 & IRAS~06361-6217 & H$_{\rm II}$ \\
41 & IRAS~07145-2914 & S2 \\
42 & IRAS~07449+3350 & H$_{\rm II}$ \\
43 & IRAS~07598+6508 & S1 \\
44 & IRAS~08208+3211 & LINER \\
45 & IRAS~08559+1053 & S2 \\ 
46 & IRAS~09022-3615 & S2 \\
47 & IRAS~09463+8141 & LINER\\
48 & IRAS~10091+4704  & LINER\\
49 & IRAS~10378+1109 & LINER \\
50 & IRAS~10565+2448 & H$_{\rm II}$\\
51 & IRAS~11038+3217 & LINER\\
52 & IRAS~11095-0238 & LINER\\
53 & IRAS~11223-1244 & LINER\\
54 & IRAS~11582+3020 & LINER\\
55 & IRAS~12018+1941 & H$_{\rm II}$\\
56 & IRAS~12032+1707 & LINER\\
57 & IRAS~12072-0444 & S2\\
58 & IRAS~12205+3356 & H$_{\rm II}$\\
59 & IRAS~12514+1027 & S2\\ 
60 & IRAS~13120-5453 & H$_{\rm II}$\\
61 & IRAS~13218+0552 & S1\\
62 & IRAS~13342+3932 & S2\\ 
63 & IRAS~13352+6402 & H$_{\rm II}$\\
64 & IRAS~13451+1232 & S2\\

\hline

\end{tabular}
\end{table}

\begin{table}
\contcaption{IDs and Classification of the ULIRGs and quasars in our sample}
\begin{tabular}{lcc} \hline
ID & Galaxy & Classification  \\
\hline
65 & IRAS~14070+0525 & LINER\\
66 & IRAS~14378-3651 & LINER\\
67 & IRAS~15001+1433 & H$_{\rm II}$\\
68 & IRAS~15206+3342 & S2 \\ 
69 & IRAS~15462-0450 & S2\\ 
70 & IRAS~16090-0139 & H$_{\rm II}$ \\
71 & IRAS~16300+1558 & LINER \\
72 & IRAS~16334+4630 & H$_{\rm II}$\\
73 & IRAS~16576+3553 & H$_{\rm II}$ \\
74 & IRAS~17068+4027 & H$_{\rm II}$ \\
75 & IRAS~17179+5444 & S2 \\
76 & IRAS~17208-0014 & LINER\\
77 & IRAS~17252+3659 & H$_{\rm II}$ \\
78 & IRAS~17463+5806 & LINER \\
79 & IRAS~18030+0705 & S2 \\
80 & IRAS~18443+7433 & H$_{\rm II}$ \\
81 & IRAS~192547245south & S2 \\
82 & IRAS~19297-0406 & LINER \\
83 & IRAS~19458+0944 & LINER \\
84 & IRAS~20037-1547 & S2 \\
85 & IRAS~20087-0308 & H$_{\rm II}$ \\
86 & IRAS~20100-4156 & LINER \\
87 & IRAS~20414-1651  & LINER \\
88 & IRAS~20551-4250 & H$_{\rm II}$ \\
89 & IRAS~21272+2514 & LINER \\
90 & IRAS~23060+0505 & S1 \\
91 & IRAS~23128-5919 & S2 \\
92 & IRAS~23129+2548 & LINER \\
93 & IRAS~23230-6926 & LINER \\
94 & IRAS~23253-5415 & LINER \\
95 & IRAS~23498+2423 & S2 \\
96 & 3C~273 & S1\\
97 & Mrk~1014 & S2 \\
98 & Mrk~463E & S1 \\
99 & PG~1119+120 & S1\\
100 & PG~1211+143 & S1\\
101 & PG~1351+640 & S1\\
102 & PG~2130+099 & S1\\
103 & PG~0052+251 & S1\\
104 & PG~0804+761 & S1\\
105 & PG~1116+215 & S1\\
106 & PG~1151+117 & S1 \\
107 & PG~1307+085 & S1  \\
108 & PG~1309+355 & S1 \\
109 & PG~1402+261 & S1 \\
110 & PG~1501+106 & S1 \\
111 & PG~1535+547 & S1 \\
112 & IZw1 & S1 \\
113 & PG~0049+171 & S1 \\
114 & PG~0921+525 & S1\\
115 & PG~0923+129 & S1\\
116 & PG~0934+013 & S1\\
117 & PG~1011-040 & S1 \\
118 & PG~1012+008 & S1 \\
119 & PG~1022+519 & S1\\
120 & PG~1048+342 & S1 \\
121 & PG~1114+445 & S1 \\
122 & PG~1115+407 & S1 \\
123 & PG~1149-110 & S1 \\
124 & PG~1202+281 & S1 \\
125 & PG~1244+026 & S1 \\
126 & PG~1310-108 & S1 \\
127 & PG~1322+659 & S1 \\
128 & PG~1341+258 & S1 \\
129 & PG~1351+236 & S1 \\
\hline

\end{tabular}
\end{table}
\begin{table}
\contcaption{IDs and Classification of the ULIRGs and quasars in our sample}
\begin{tabular}{lcc} \hline
ID & Galaxy & Classification  \\
\hline
130 & PG~1404+226 & S1 \\
131 & PG~1415+451 & S1 \\
132 & PG~1416-129 & S1 \\
133 & PG~1448+273 & S1 \\
134 & PG~1519+226 & S1 \\
135 & PG~1534+580 & S1 \\
136 & PG~1552+085 & S1 \\
137 & PG~1612+261 & S1 \\
138 & PG~2209+184 & S1 \\
139 & PG~2304+042 & S1 \\
\hline
\end{tabular}
\end{table}
}